\documentstyle[12pt]{article}

\newcommand{\bcal}[1]{{\mbox{\boldmath $\cal #1$}}}
\newcommand{\bgr}[1]{{\mbox{\boldmath $#1$}}}

\date{February 1994\\DOE-ER\,40757-040 /
CPP-93-7 / LTP-041-UPR}
\title{The zero-momentum limit of thermal green functions}

\author{Jos\'e F. Nieves\\
Department of Physics, University of Puerto Rico\\
P. O. Box 23343, Rio Piedras, Puerto Rico 00931\and
Palash B. Pal\\
Center for Particle Physics\\
University of Texas Austin, TX 78712, USA
}

\begin{document}

\maketitle

\begin{abstract}

The zero momentum limit of thermal self-energies calculated in
perturbation theory depends on the order in which the time and the
space components of the momentum are taken to zero. We show that this
is an artifact of the perturbative calculation, and in fact the limit
is well-defined when higher orders of the perturbation expansion are
properly taken into account.

\end{abstract}

The existing calculations of thermal self-energy functions using
the formalism of Thermal Field Theory
yield results that are not defined if the external
momentum 4-vector is zero.  The classic example is
the photon self-energy $\pi_{\mu\nu}$ in an electron gas. It is well
known \cite{wellknown} that the result of the one-loop calculation
of $\pi_{\mu\nu}(p^0,\vec p)$ for a photon with external momentum
$p^\mu = (p^0,\vec p)$ is such that
\begin{eqnarray}
\lim_{| \vec p | \to 0} \pi_{\mu\nu} (0, \vec p) \neq
\lim_{p^0 \to 0} \pi_{\mu\nu} (p^0, \vec 0)\,,
\label{problem}
\end{eqnarray}
so that the limit in which all components go to zero is not defined.
To be more specific, the above-mentioned problem
occurs for the real part of the self-energy, while
the imaginary part is well defined.

Various attempts have been made to resolve this puzzle
\cite{MNU85,FuYa88,Eva88,GrHo90}, which involve either introducing new
and ad-hoc Feynman rules for thermal field theories, or putting
restrictions on the general rules.  Along another line of attempt, it has
recently been pointed out by Arnold, Vokos, Bedaque and Das (AVBD)
\cite{AVBD93} that the problem mentioned above occurs only if the self
energy diagram contains two propagators of the same mass. If the masses of
the particles in the loop are different, the problem does not exist.
In fact, the calculations of the neutrino self-energy in a gas of
electrons and nucleons, which were carried out even before the work of
Ref. \cite{AVBD93}, show this \cite{RaNo88,PaPh89,Nie89}.  It has been
speculated that this property may be utilized to introduce
a mass-splitting regularization for thermal diagrams \cite{AVBD93} in
cases where problems are known to occur.

This problem, as well the attempts to resolve it, are
based on the results of one-loop perturbative calculations.
It is natural to ask whether the singular behaviour of the self-energy
function at zero momentum might be a consequence of the approximations
and idealizations that are implicitly made in the perturbative
calculations. In this article we show that this is precisely the case,
and that the calculation of the self-energy beyond one-loop order
yields a result that is defined at zero momentum. This result is
obtained by calculating the self-energies using the full propagators
for the particles that appear in the internal lines of the loop
diagrams. Some of these propagators have an absorptive part which, as
observed by AVBD \cite{AVBD93}, is well-defined at zero momentum even
if they are calculated to one-loop and the particles in the internal
lines of the loop have the same mass. As we will see, this in turn
governs the zero-momentum limit of those self-energy diagrams in which
the internal lines have the same mass.  We exemplify these assertions
by calculating the photon self-energy in scalar QED, but similar
considerations apply to QED proper as well.

Before proceeding, we recapitulate some results of the
canonical approach to the thermal propagators,
which we will be using
throughout \cite{Nie90}. In this approach, one has to use
anti-time-ordered propagators in addition to the time-ordered
ones, as well as propagators with no time-ordering.
These propagators
can be arranged in the form of a $2\times 2$ matrix. For example, for
any bosonic field $\Phi^A$ where $A$ denotes any Lorentz index
carried by the field (none for a scalar field, one for the photon), we
can write
        \begin{eqnarray}
i \bcal D_{11}^{AB} (x-y) &\equiv &
\left< {\cal T} \; \Phi^A (x) \Phi^{B\ast} (y) \right> \,, \\*
i \bcal D_{22}^{AB} (x-y) &\equiv &
\left< \overline{{\cal T}} \; \Phi^A (x) \Phi^{B\ast} (y) \right>
\,, \\*
i \bcal D_{12}^{AB} (x-y) &\equiv & \left<
\Phi^{B\ast} (y) \Phi^A (x) \right> \,, \\*
i \bcal D_{21}^{AB} (x-y) &\equiv &
\left< \Phi^A (x) \Phi^{B\ast} (y) \right> \,,
        \end{eqnarray}
where the time-ordering and the anti-time-ordering operators $\cal T$ and
$\overline{{\cal T}}$ are defined as
        \begin{eqnarray}
{\cal T} \; \Phi^A (x) \Phi^{B\ast} (y)
&\equiv& \Theta(x_0-y_0) \Phi^A (x) \Phi^{B\ast} (y) +
\Theta (y_0-x_0) \Phi^{B\ast} (y) \Phi^A (x) \,,\\*
\overline{{\cal T}} \; \Phi^A (x) \Phi^{B\ast} (y)
&\equiv& \Theta(y_0-x_0) \Phi^A (x) \Phi^{B\ast} (y) +
\Theta (x_0-y_0) \Phi^{B\ast} (y) \Phi^A (x) \,,
        \end{eqnarray}
$\Theta$ being the step function.  We can now make the momentum
space expansion of the field as
        \begin{eqnarray}
\Phi^A (x) = \int {d^3p \over (2\pi)^3 2E} \sum_\lambda \left[
a_\lambda(p) u^A(p,\lambda) e^{-ip \cdot x} +
b^\ast_\lambda(p) v^A(p,\lambda) e^{ip \cdot x} \right] \,,
        \end{eqnarray}
where $u^A$ and $v^A$ represent different plane wave solutions
arranged by the index $\lambda$, and $a_\lambda(p)$ and
$b_\lambda(p)$ are the annihilation operators for particles and
antiparticles, respectively. For a self-adjoint field like the photon,
$a_\lambda(p) = b_\lambda(p)$.
The properties of  the thermal
bath come in from the expectation values
        \begin{eqnarray}
\left< a_\lambda(p) a^\ast_{\lambda'} (p') \right> &=& (2\pi)^3 2E \delta
(\vec p - \vec p\,') \delta_{\lambda\lambda'} \left[ 1 +  f_B (p, \alpha)
\right] \,, \\*
\left< b_\lambda(p) b^\ast_{\lambda'} (p') \right> &=& (2\pi)^3 2E \delta
(\vec p - \vec p\,') \delta_{\lambda\lambda'} \left[ 1 +  f_B (p,
-\alpha) \right] \,,
        \end{eqnarray}
with
        \begin{eqnarray}
f_B (p, \alpha) = {1 \over e^{\beta p\cdot u - \alpha} - 1} \,,
\label{f}
        \end{eqnarray}
where $\alpha$ plays the role of a chemical potential.
We have introduced the velocity 4-vector $u^\mu$ of the
heat bath, which has components $(1,\vec 0)$ in its own
rest frame.

For scalar fields, the procedure described above gives \cite{Nie90} the
$2\times 2$ free-field propagator as
        \begin{eqnarray}
\bgr \Delta (p) = \bcal U_B
\left( \begin{array}{cc} \Delta_0  & 0 \\ 0 & -\Delta_0^*
\end{array} \right)  \bcal U_B \,,
\label{boseprop}
        \end{eqnarray}
where
        \begin{eqnarray}
\Delta_0 \equiv {1 \over p^2 - m^2 + i0} \,,
        \end{eqnarray}
and
        \begin{eqnarray}
\bcal U_B = {1 \over \sqrt{1 + \eta_B(p,\alpha)}}
\left( \begin{array}{cc} 1 +  \eta_B(p,\alpha) &
\epsilon(p\cdot u) f_B(p,\alpha) \\
-\epsilon(p\cdot u) f_B(-p,-\alpha) &
1 +  \eta_B(p,\alpha) \end{array} \right) \,,
        \end{eqnarray}
with $\epsilon(x) \equiv \Theta(x) - \Theta(-x)$, and
        \begin{eqnarray}\label{defeta}
\eta_B(p,\alpha) = \Theta(p\cdot u) f_B (p,\alpha) + \Theta(-p\cdot u)
f_B (-p,-\alpha) \,.
        \end{eqnarray}
It is straightforward to check that this gives, for example,
        \begin{eqnarray}
{\bgr \Delta}_{11} (p) & = & \frac{1}{p^2 - m^2 + i0} -
2\pi i \delta (p^2-m^2) \eta_B(p,\alpha)\,,\label{Delta11}
        \end{eqnarray}
which is the propagator given by Dolan and Jackiw~\cite{DoJa74}.
The explicit forms of the other components are also given in the
literature~\cite{others}.

The one-loop diagrams for the photon self-energy in a background
of $\phi$ particles are depicted in Fig.~\ref{f:gam}.
Diagram~\ref{f:gam}b
produces a term that is proportional to the total
electric charge of the system.  However, since that
term is a constant, independent of the photon momentum,
it will not be relevant for our discussion and we
will not consider it any further.

The result of calculating Diagram~\ref{f:gam}a using the free-field
propagator given above for the $\phi$ field is
\begin{eqnarray}\label{pisingular}
\mbox{Re} \, [\bgr \pi_{\mu\nu}(k_0, \vec k)]_{11} & = &
e^2 \int \frac{d^4p}{(2\pi)^3} \eta_B(p,\alpha) \delta(p^2 - m^2)\nonumber\\
& & \quad\times\left\{\frac{(2p + k)_\mu (2p + k)_\nu}{k^2 + 2p\cdot k}
+ (k\rightarrow -k)\right\}
\end{eqnarray}
where the vacuum contribution has been omitted,
as we will always do henceforth
whenever we
write explicit expressions for the self-energies.
The above formula reveals the problem to which we alluded
in Eq.\ (\ref{problem}).
As we mentioned earlier, this problem vanishes if the diagram
is evaluated employing the full propagator of the
$\phi$ field instead of the free-field propagator.
The full scalar propagator, which we denote
by ${\bgr\Delta}'(p)$,
can be written just like in Eq.\ (\ref{boseprop})
but with $\Delta_0$ replaced by
        \begin{eqnarray}
\Delta'_0 =  {1 \over p^2 - m^2 - \Pi_0},
        \end{eqnarray}
where $\Pi_0$ is the self-energy function for the $\phi$ field.
Thus,
        \begin{eqnarray}
\bgr \Delta'(p) = \bcal U_B
\left( \begin{array}{cc} \Delta'_0  & 0 \\ 0 & -\Delta_0^{\prime\ast}
\end{array} \right)  \bcal U_B \,,
\label{bosepropexact}
        \end{eqnarray}
and in particular,
        \begin{eqnarray}
{\bgr\Delta}'_{11} (p) =  \Delta'_0 + \left[ \Delta'_0 -
\Delta^{\prime\ast}_0
\right]  \eta_B(p,\alpha) \,.
\label{Delta'11}
        \end{eqnarray}

The important consequence of replacing $\Delta_0$ by $\Delta'_0$
is that the $\delta$-function present in Eq.\ (\ref{Delta11}) is now
smeared if the absorptive part of $\Pi_0$ is non-zero.  For this
reason it is the easy to see that,
if Diagram~\ref{f:gam}a is calculated with the propagator
${\bgr\Delta}'_{11}$ for the $\phi$ field instead of
the free particle propagator,
then the problem at zero momentum does not arise in evaluating
$[\bgr \pi_{\mu\nu}(k)]_{11}$.
Notice that the dispersive part of $\Pi_0$ plays no role
in this argument.  It is not difficult
to see that retaining only the dispersive part of $\Pi_0$
and neglecting its absorptive part does not remove
the singularity of $[\bgr \pi_{\mu\nu}]_{11}$ at zero momentum.

The next step is to calculate $\Pi_0$ and show that in general
it has an absorptive part.
To this end, we recall that the inverse of the full scalar
propagator is given by
        \begin{eqnarray}\label{bospropinv}
{\bgr \Delta}'^{-1} (p) = p^2 - m^2 -\bgr\Pi\,,
        \end{eqnarray}
where $\bgr\Pi$ is a $2\times 2$ matrix whose components
must be calculated using the Feynman rules of the theory.
Comparing Eqs.\ (\ref{bosepropexact}) and (\ref{bospropinv}), the
following relations are obtained,
        \begin{eqnarray}
\bgr \Pi_{11} &=& \Pi_0 + (\Pi_0 - \Pi_0^\ast )
\eta_B (p,\alpha) \\
\bgr \Pi_{22} &=& -\Pi_0^\ast + (\Pi_0 - \Pi_0^\ast )
\eta_B (p,\alpha) \\
\bgr \Pi_{12} &=& -(\Pi_0 - \Pi_0^\ast )
\epsilon (p\cdot u) f_B (p,\alpha)
\label{pi12}\\
\bgr \Pi_{21} &=& -(\Pi_0 - \Pi_0^\ast )
\epsilon (-p\cdot u) f_B (-p,-\alpha)
\label{pi21}\,.
        \end{eqnarray}
from which it is easily seen that
        \begin{eqnarray}\label{realparteq}
\mbox{Re}\, \Pi_0 (p) &=& \mbox{Re}\, \bgr \Pi_{11}(p) \\
\mbox{Im}\, \Pi_0 (p) &=& {\epsilon(p\cdot u) \bgr \Pi_{12}(p) \over 2i
f_B(p,\alpha)} \,.\label{absparteq}
        \end{eqnarray}
Therefore, to determine $\Pi_0$ we must calculate $\bgr \Pi_{11}$ and
$\bgr \Pi_{12}$, which can be done by evaluating the diagrams in
Fig.~\ref{f:phi}. Since the dispersive part $\Pi_0$ is not relevant for
resolving the zero momentum problem of the photon
self-energy, we will not calculate it here. However,
for the consistency of our scheme it is important to stress that
since the internal lines in Fig.~\ref{f:phi} correspond to particles
of different mass, the function $\mbox{Re}\,\bgr\Pi_{11}$ (and hence
$\mbox{Re}\,\Pi_0$) does not suffer from the zero momentum problem according
to the observation of AVBD~\cite{AVBD93}.

We now turn the attention to the calculation of the absorptive
part of $\Pi_0$.  The simplest way to proceed is to
calculate
$\bgr \Pi_{12}$ and then use Eq.\ (\ref{absparteq}). As in the case
of Diagram \ref{f:gam}b for the photon
self-energy, the diagram in Fig.~\ref{f:phi}b is irrelevant for our
purpose. In Fig.~\ref{f:phi}a, the
scalar propagator that enters for
$\bgr \Pi_{12}$ is obtained from Eq.\ (\ref{boseprop}) as
			\begin{equation}
\bgr\Delta_{12}(p') = -2\pi i\delta(p^{\prime 2} - m^2)f_B(p',\alpha)
\epsilon(p'\cdot u)\,.
\label{Delta12}
			\end{equation}
In addition to this, we also need the thermal photon propagator, whose
form depends on the gauge that is chosen. Here we will use the Coulomb
gauge which, for several reasons, has been advocated as a convenient
one for carrying out finite temperature calculations in QED
\cite{DolNiev1,DolNiev2}.  The only component that is needed for the
calculation at hand is the 21 component which, borrowing from Refs.
\cite{DolNiev1,DolNiev2}, is given by
\begin{equation}\label{photonprop}
\bgr D_{21}^{\mu\nu}(k) = 2\pi i\delta(k^2)f_B(-k,0)
\epsilon(k\cdot u)(-S^{\mu\nu})\,,
\end{equation}
where
\begin{equation}\label{Smunu1}
S_{\mu\nu} = g_{\mu\nu} + \frac{1}{\kappa^2}k_\mu k_\nu -
\frac{\omega}{\kappa^2}(u_\mu k_\nu + k_\mu u_\nu)\,.
\end{equation}
Here $\omega$ and $\kappa$ are defined by
\begin{eqnarray}
\omega  =  k\cdot u\,, \qquad
\kappa  =  \sqrt{\omega^2 - k^2}\,,
\end{eqnarray}
and they represent the energy and momentum of the photon
in the rest frame of the heat bath.
A particularly useful relation is
\begin{equation}\label{Smunu2}
\left. \sum_{\lambda = 1,2} \epsilon_\mu(k,\lambda)
\epsilon_\nu(k,\lambda) \right|_{\omega = \kappa} =
- \left. \vphantom{\sum_a 1} S_{\mu\nu}\right|_{\omega = \kappa}\,,
\end{equation}
where the polarization vectors $\epsilon_\mu(k,\lambda) = (0,\vec
e(k,\lambda))$ are such that, in the rest frame of the heat bath,
\begin{equation}
\vec k \cdot \vec e(k,1) = \vec k \cdot \vec e(k,2) = 0 \,.
\end{equation}

The application of the Feynman rules to the diagram of
Fig.~\ref{f:phi}a gives
\begin{eqnarray}
-i\bgr\Pi_{12}(p) = (-ie)(ie)\int\frac{d^4p'}{(2\pi)^4}
i\bgr D^{\mu\nu}_{21}(k)(p + p')_\nu
i\bgr\Delta_{12}(p')(p + p')_\mu\,,
\end{eqnarray}
where we have defined
\begin{equation}
k = p' - p \,.
\end{equation}
Substituting the photon and scalar propagators
into this expression and using Eq.\ (\ref{absparteq})
we obtain
\begin{eqnarray}\label{impi0}
\mbox{Im}\, \Pi_0(p) & = & -\frac{e^2}{2} (2\pi)^2 \epsilon(p\cdot u)
\int\frac{d^4 p'}{(2\pi)^4}\delta(k^2)\delta(p^{\prime 2} - m^2)
\epsilon(p'\cdot u)\epsilon(k\cdot u) \nonumber\\
& & \times
(-S^{\mu\nu})(p + p')_\mu(p + p')_\nu
(f_B(k,0) - f_B(p',\alpha))\,,
\end{eqnarray}
where the identity
\begin{equation}
f_B(-k,0)f_B(p',\alpha)
= f_B(p,\alpha)[f_B(p',\alpha) - f_B (k,0)]
\end{equation}
has been used.  Eq.\ (\ref{impi0}) can be written in the form
\begin{eqnarray}
\mbox{Im}\, \Pi_0(p) & = & -\frac{e^2}{2}\epsilon(p\cdot u) \int
\frac{d^3 p'}{(2\pi)^3 2E'}
\frac{d^3 k}{(2\pi)^3 2\omega}(2\pi)^4(-S^{\mu\nu})\nonumber\\*
& & \mbox{}\times\left\{\right.
\delta^{(4)}(p + k - p')(p + p')_\mu(p + p')_\nu
[f_\gamma - f_\phi]\nonumber\\*
& & \quad + \delta^{(4)}(p - k + p')(p - p')_\mu (p -
p')_\nu
[\overline f_\phi - f_\gamma]\nonumber\\*
& & \quad + \delta^{(4)}(p - k - p')(p + p')_\mu(p + p')_\nu
[1 + f_\gamma + f_\phi]\nonumber\\*
& & \quad + \delta^{(4)}(p + k + p')(p - p')_\mu (p - p')_\nu
[1 + f_\gamma + \overline f_\phi]
\left.\right\}\,,
\end{eqnarray}
where, for the sake of brevity, we have called the photon and $\phi$
particle distributions by
\begin{eqnarray}
f_\gamma & = & f_B (k,0) \,, \nonumber\\*
f_\phi & = & f_B (p',\alpha) \,,
\end{eqnarray}
while $\overline f_\phi$ is the antiparticle density distribution,
which is obtained from $f_\phi$ by changing the sign of $\alpha$.
In addition,
\begin{eqnarray}
p^{\prime\mu} = (E',\vec p^{\,\prime})\,, &\qquad &
E' = \sqrt{\vec p^{\,\prime 2} + m^2}\,, \\*
 k^\mu = (\omega,\vec k)\,, &\qquad & \omega = |\vec k|\,.
\end{eqnarray}
A more physically intuitive representation of this formula
can be obtained by using Eq.\ (\ref{Smunu2})
for $S_{\mu\nu}$.  Then, introducing the amplitudes
\begin{eqnarray}
M_A & = & (-ie)(p + p')^\mu\epsilon_\mu(k,\lambda)\,,\\*
M_B & = & (-ie)(p - p')^\mu\epsilon_\mu(k,\lambda)\,,
\end{eqnarray}
we get
\begin{equation}
\mbox{Im}\, \Pi_0 = - \left| p\cdot u \right| \Gamma(p)\,,
\end{equation}
where we have defined
\begin{eqnarray}\label{totalrate}
\Gamma(p) & \equiv & \frac{1}{2p\cdot u}
\int{\frac{d^3k}{(2\pi)^3 2\omega}\frac{d^3p'}{(2\pi)^3
2E'}}(2\pi)^4 \nonumber\\*
& & \mbox{}\times\left\{\right.\delta^{(4)}(p + k -
p')[f_\gamma(1 + f_\phi) - f_\phi(1 + f_\gamma)]{\sum_{\lambda =
1,2}|M_A|^2}\nonumber\\*
& & \quad + \delta^{(4)}(p - k + p')[\overline f_\phi(1 +
f_\gamma) - f_\gamma(1 + \overline f_\phi)]{\sum_{\lambda =
1,2}|M_B|^2}\nonumber\\*
& & \quad + \delta^{(4)}(p - k - p')[(1 + f_\gamma)(1 +
f_\phi) - f_\gamma f_\phi]{\sum_{\lambda = 1,2}|M_A|^2}\nonumber\\*
& & \quad + \delta^{(4)}(p + k + p')[f_\gamma \overline f_\phi
- (1 + f_\gamma)(1 + \overline f_\phi)]{\sum_{\lambda = 1,2}|M_B|^2}
\left.\right\}\,.
\end{eqnarray}
The formula for $\Gamma$ given in Eq.\ (\ref{totalrate})
is immediately recognized
as the total rate for a $\phi$ particle of energy $p^0$ and
momentum $\vec p$ (as seen from the rest frame of the medium)
with integrations over the phase space weighted by
the statistical factors appropriate for each process \cite{Wel83}.
$M_A$ is the
amplitude for $\gamma\phi\rightarrow \phi$ or the decay
$\phi\rightarrow \gamma\phi$, while $M_B$ is the amplitude
for $\phi\overline\phi \rightarrow\gamma$ or
$\gamma\phi\overline\phi\rightarrow 0$.
The amplitudes for the inverse reactions are given by the
complex conjugates of $M_A$ and $M_B$.  For certain specific values
of $p^0$ and $\vec p$ some of these processes will
be kinematically forbidden, but in general $\Gamma$ is non-zero.

In conclusion,
the one-loop photon self-energy calculated with the full
$\phi$ propagator given in Eq.\ (\ref{Delta'11})
instead of the free propagator of Eq.\ (\ref{Delta11}),
is defined at zero momentum.  In particular, the absorptive
part of the $\phi$ self-energy, which physically
is related to the damping rate of the particle,
cannot be neglected if the photon self-energy
is evaluated at zero momentum.  Then, the physical
picture that emerges is the following.
The traditional formulas that are given for
\begin{eqnarray}
\lim_{| \vec p | \to 0} \pi_{\mu\nu} (0, \vec p)
\end{eqnarray}
and
\begin{eqnarray}
\lim_{p^0 \to 0} \pi_{\mu\nu} (p^0, \vec 0)\,,
\end{eqnarray}
which are related to well known physical quantities
such as the plasma frequency and Debye radius,
are valid in the limiting cases
\begin{eqnarray}
p^0 = 0\,&;& \Gamma \ll| \vec p| \ll m\nonumber\\
\vec p = 0\,&;& \Gamma \ll p^0 \ll m\,.
\end{eqnarray}
Since the two limits correspond to two different
physical situations the results are different.
Traditionally $\Gamma$ is omitted in the above conditions,
but then it must be kept in mind that the formulas
cannot be taken literally all the way to zero momentum. The same
conclusion can be reached for fermionic QED and other field theories,
which will be discussed in detail in a future publication.
\newpage

\newpage
\centerline{\bf Figure captions}

\begin{enumerate}
\item{Photon self-energy diagrams.\label{f:gam}}
\item{Self-energy diagrams for the scalar.\label{f:phi}}
\end{enumerate}

\end{document}